\documentclass[journal]{IEEEtran}
\usepackage{amsmath}
\usepackage{stmaryrd}
\usepackage{amsfonts}
\usepackage{mathrsfs}
\usepackage{amssymb,color,balance,cite}
\usepackage[dvips]{graphicx}
\usepackage{amsmath,cite,color,amssymb,cite,color}
\usepackage{verbatim}

\makeatletter

\newcommand{\Rmnum}[1]{\expandafter\@slowromancap\romannumeral #1@}
\makeatother

 \allowdisplaybreaks[4]

\begin{document}

\title{ Radiation Efficiency Aware High-Mobility Massive MIMO with Antenna Selection}
\author{Zhinan Hu and  Weile Zhang
\thanks{
Z. Hu and W. Zhang are with the MOE Key Lab for Intelligent Networks and Network Security, Xi'an Jiaotong University, Xi'an, Shaanxi, 710049, China. (email: wlzhang@mail.xjtu.edu.cn)
 }
}

 \maketitle

\vspace{-15mm}

\begin{abstract}
High-mobility wireless communications have received
a lot of attentions in the past few years. In this paper,
we consider angle-domain Doppler shifts compensation scheme
to reduce the channel time variation for the high-mobility uplink
from a high-speed terminal (HST) to a base station (BS). We
propose to minimize Doppler spread by antenna weighting under
the constraint of maintaining radiation efficiency. The sequential
parametric convex approximation (SPCA) algorithm is exploited
to solve the above non-convex problem. Moreover, in order to save the
RF chains, we further exploit the idea of antenna selection for
high-mobility wireless communications. Simulations verify the
effectiveness of the proposed studies.
\end{abstract}

\IEEEpeerreviewmaketitle

\section{Introduction}

Over the past few years, the research on high-mobility wireless communication has received a lot of attentions~\cite{Wu16}. The multiple Doppler shifts caused by relative motion between transceivers superimpose at the receiver, resulting in fast time fluctuations of the channel and bringing severe inter-carrier interference (ICI).
The approach to estimate and compensate the Doppler shifts
from spatial domain has been widely studied.
For sparse high-mobility communication scenarios with a few dominating propagation paths, small-scale uniform circular antenna array (UCA) and uniform linear antenna array (ULA) can be adopted to separate the multiple Doppler shifts and eliminate ICI via array beamforming~\cite{Zhang11}.

More recently, the large-scale antenna array or massive multi-input multi-output (MIMO) has attracted increasing interests in dealing with richly scattered high-mobility scenarios due to its high spatial resolution~\cite{Marzetta10,Larsson14,Zhang18}. For high-mobility downlink, the work in~\cite{downGuo17} separated the Doppler shifts by beamforming network with a large-scale antenna array and proposed a joint oscillator frequency offset and Doppler shift estimation scheme. For uplink transmissions, it has been demonstrated that the harmful effect of Doppler shift can be also mitigated by massive MIMO technique~\cite{upGuo17,Ge18}.
For example, the authors in~\cite{upGuo17} proposed a transmit multi-branch beamforming scheme where the angle domain Doppler shifts compensation is exploited to suppress the uplink channel variation.
The corresponding exact power spectrum density (PSD) of uplink equivalent channel is derived in~\cite{Ge18}, and an antenna weighting technique is further proposed to reduce the channel time variation. However, the antenna weighting scheme in~\cite{Ge18} only consider the target of Doppler spread reduction and may lead to a severe loss of energy radiation efficiency. This motivates our current work.

In this paper, we consider angle-domain Doppler shifts compensation scheme to reduce the channel time variation for
the high-mobility uplink from a high-speed terminal (HST) to
a base station (BS). We formulate Doppler spread minimization
problem by antenna weighting under the constraint of maintaining radiation efficiency. We then exploit the sequential parametric convex approximation (SPCA) algorithm to solve the above non-convex problem. On the other side, in view
of the fact that the fully connected massive MIMO may
suffer from prohibited high cost and complexity~\cite{Nai10}, we further exploit the idea of antenna selection for high-mobility wireless communications. Simulation results are provided to verify the proposed studies.

\section{System Model}

\begin{figure}[t]
\vspace{-5mm}
\centering
\includegraphics[width=9cm]{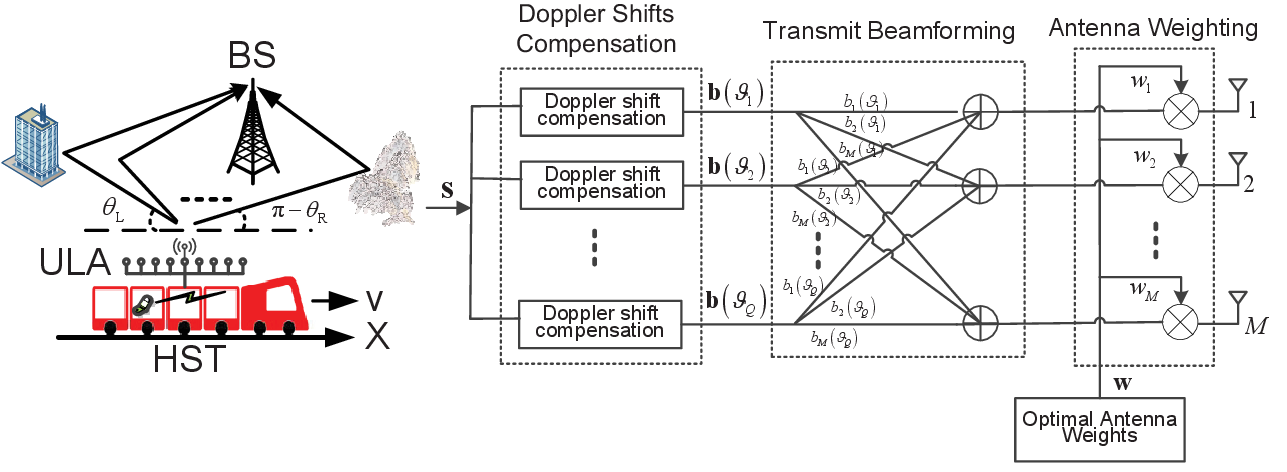}
\vspace{-5mm}
\caption{Uplink transmission with angle-domain Doppler shifts compensation from a HST to a BS.}\label{fig:insert}
\end{figure}

As illustrated in Fig. 1, we consider the high-mobility uplink transmission from a HST to BS, where an $M$-element ULA is equipped at HST. Denote the normalized antenna space of ULA as $d$, which is the ratio between the antenna space and the radio wavelength. Let the direction of ULA coincide with that of HST motion.
We consider the flat fading channel for simplicity. Nevertheless, the proposed scheme can be directly applied to frequency selective channels.
Specifically, the channel from HST to BS is composed of a bunch of propagation paths. We assume the angle of departure (AOD) of these paths is constrained within a region of $(\theta_L, \theta_R)$.

Denote ${{\bf s}} = [{s}(0),{s}(1),...,{s}(N - 1)]^T$ as the length-$N$ transmitted time domain symbols in one orthogonal frequency division multiplexing (OFDM) block. We consider the antenna weighting technique~\cite{Ge18} is adopted, where the $m\rm{th}$ antenna is weighted by a complex weight $w_m$.
 The multi-branch transmit matched filtering (MF) beamforming is performed towards a set of $Q$ selected directions
${{\vartheta }_{q}}\in({{\theta }_{L}},{{\theta }_{R}}),q=1,2,...,Q$.
The $q$th MF beamformer is ${\bf b}({\vartheta _q}) = \zeta \cdot{\rm{\bf a}(}{\vartheta _q}){e^{j\pi \phi_q }}$, where ${\bf a}(\vartheta_q) = [1, {\rm e}^{j2\pi d \cos\vartheta_q}, \cdots, {\rm e}^{j2\pi d (M-1)\cos\vartheta_q}]^T$ denotes the array response vector corresponding to direction $\vartheta_q$,  $\zeta$ is the normalization coefficient to keep the total transmit power per symbol to 1 and $\phi_q $ denotes the introduced random phase.

Let ${\bf\Phi}(  \epsilon ) = \textrm{diag}(1, {\rm e}^{  j \omega_d  T_s \epsilon }, \cdots, {\rm e}^{  j \omega_d T_s (N-1)\epsilon } )$ represent the $N\times N$ diagonal phase rotation matrix introduced by frequency shift of $ f_d \epsilon$,  with $\omega_d = 2\pi f_d$  and $T_s$ being sampling interval.
After transmit beamforming and Doppler shift compensation, the transmitted $M\times N$ signal matrix at transmit antenna array can be expressed as $ \textrm{diag}( {\bf w}) {\bf b}^*({\vartheta _q}) {\bf s}^T {\bf\Phi}(  -\cos\vartheta_q  )  $. ${\bf w} = [w_1, w_2, \cdots, w_M]^T$ denotes the weighting vector.
Then, according to~\cite{Ge18}, the received signal at BS can be expressed in the form of
\begin{align}
 {\bf r} = & \sum_{q=1}^Q  \int_{\theta_L}^{\theta_R}  \alpha(\theta)
{\bf a}(\theta)^T \textrm{diag}( {\bf w})  {\bf b}^*({\vartheta _q}) {\bf s}^T \nonumber \\ & \kern 80pt \times {\bf\Phi}(  -\cos\vartheta_q  ) {\bf\Phi}(  \cos\theta  )    d\theta,
\end{align}
where $\alpha(\theta)$ represents the complex gain of the path associated with AOD $\theta$. Then,  the equivalent uplink channel can be expressed as
\begin{align}
& g(t) =  \zeta \cdot \sum_{q=1}^Q  \int_{\theta_L}^{\theta_R} \alpha(\theta) G(\cos\theta, \cos\vartheta_q)  \\ & \kern 60pt
{\rm e}^{j \omega_d (\cos\theta - \cos\vartheta_q) t + j\phi_q} d\theta \nonumber
\end{align}
where $G(\cos\theta, \cos\vartheta_q) = \frac{1}{M} {\bf a}^H(\vartheta_q) \textrm{diag}({\bf w}) {\bf a}(\theta)$.

Assume that the directions are selected according to `Equi-cos' criterion in~\cite{Ge18} such that $\cos {\vartheta _q},q = 1,2,...,Q$ are evenly distributed between $(\cos \theta_R,\cos \theta_L)$.
According to~\cite{Ge18}, the channel PSD at BS can be expressed as
$ P(\omega ) = \frac{1}{{{\omega _d}}}|{\mathcal{G}}(\tilde \omega ){|^2}{W}(\tilde \omega )$,
where $\omega _d=2\pi f_d$ and $\tilde \omega={\frac{\omega }{{{\omega _d}}}}$ denotes the normalized Doppler frequency with respect to maximum Doppler shift.  Denote  ${\boldsymbol{\varsigma}}(\tilde \omega) = [1, {\rm e}^{-j2\pi d \tilde\omega}, \cdots, {\rm e}^{-j2\pi d (M-1)\tilde\omega} ]^T$.
Here, $|{\mathcal{G}}(\tilde \omega ){|^2} $ and $W(\omega)$
are respectively named as beam function and window function,
where $|{\mathcal{G}}(\tilde \omega ){|^2} = |\frac{1}{M}{\boldsymbol{\varsigma}}^T(\tilde \omega){\bf w} {|^2}$ and
 \begin{align*}
 & W(\omega )=\left\{ \begin{matrix}
   \frac{2\pi }{({{\theta }_{R}}-{{\theta }_{L}})}\frac{\arccos (\cos{{\theta }_{R}}-\tilde{\omega })-{{\theta }_{L}}}{\mu ({{\theta }_{L}},{{\theta }_{R}})},-\mu ({{\theta }_{L}},{{\theta }_{R}})\le \tilde{\omega }<0  \\
   \frac{2\pi }{({{\theta }_{R}}-{{\theta }_{L}})}\frac{{{\theta }_{R}}-\arccos(\cos{{\theta }_{L}}-\tilde{\omega })}{\mu ({{\theta }_{L}},{{\theta }_{R}})},0\le \tilde{\omega }\le \mu ({{\theta }_{L}},{{\theta }_{R}})  \\
\end{matrix} \right.
\end{align*}

Afterwards, the Doppler spread with antenna weighting can be calculated as
\begin{align}
\sigma_D = {\omega _d}\sqrt {\frac{{{{\bf w}^H}{{\bf C}_2}{\bf w}}}{{{{\bf w}^H}{{\bf C}_0}{\bf w}}}},
\end{align}
where ${{\bf C}_{0}} = \int_{ - \mu({\theta _L},{\theta _R})}^{\mu({\theta _L},{\theta _R})} {{ W}(\tilde \omega ){\boldsymbol \varsigma}^*(\tilde \omega ){\boldsymbol \varsigma}^T(\tilde \omega )d\tilde \omega }$,
${{\bf C}_{2}} = \int_{ - \mu({\theta _L},{\theta _R})}^{\mu({\theta _L},{\theta _R})} {{{\tilde \omega }^2}} { W}(\tilde \omega ){\boldsymbol \varsigma}^*(\tilde \omega ){\boldsymbol \varsigma}^T(\tilde \omega )d\tilde \omega$ and $\mu({\theta _L},{\theta _R})=\mathrm{cos}\theta _L-\mathrm{cos}\theta _R$.

Then, the normalized Doppler spread can be defined as
\begin{align}
\frac{\sigma_D}{\omega_d} =\sqrt {\frac{{{{\bf w}^H}{{\bf C}_2}{\bf w}}}{{{{\bf w}^H}{{\bf C}_0}{\bf w}}}},
\end{align}
which represents the magnitude of Doppler spread suppression benefiting from the angle-domain Doppler shifts compensation scheme.

\section{Radiation Efficiency Aware Doppler Spread Suppresion}

\begin{figure}[t]
\vspace{-5mm}
\begin{center}
\includegraphics[width=65mm]{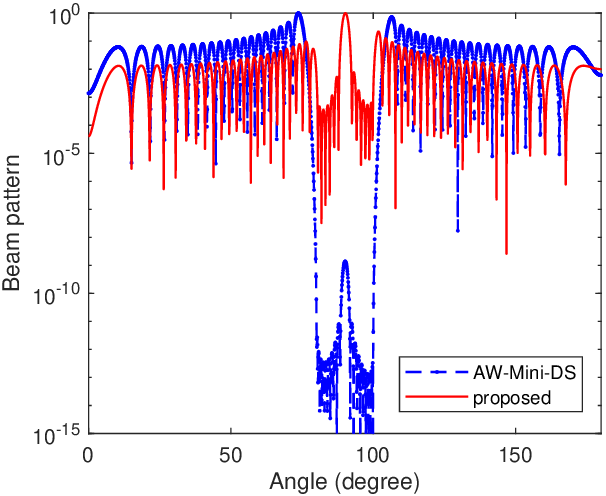}
\vspace{-5mm}
\end{center}
\caption{Comparison of the radiation pattern  between AW-Mini-DS~\cite{Ge18} and the proposed  scheme. }
\vspace{-5mm}
\end{figure}

In the scheme in~\cite{Ge18} (labelled as `AW-Mini-DS'), the antenna weighting is designed to only minimize the Doppler spread, which may lead to severe loss of transmission efficiency. Let us consider the following example. The AoDs are constrained within ($\theta_L,\theta_R$) with $\theta_L=85^{\circ},\theta_R=95^{\circ}$.
  The ULA consists of $M=64$ antennas, the maximum Doppler shift is set as $f_d=5000$Hz, and the antenna spacing is taken as $d=0.45$.
We depict the radiation pattern of the antenna array with the weighting technique in AW-Mini-DS in Fig. 2 by the solid line.  It is evident that, in this case, although AW-Mini-DS can achieve the lowest Doppler spread, i.e., the smallest channel fluctuations, its radiation mainlobe is extremely lower than the sidelobes.  This intuitively indicates that the vast majority of  radiation power in AW-Mini-DS has leak out from the expected AoD region $(85^{\circ}, 95^{\circ})$.

More mathematically, it can be observed from~\cite{Ge18} that, the total transmit signal power from antenna array can be expressed as
${{\bf w}^H}{\bf w}$, while ${{\bf w}^H}{\bf C}_0{\bf w}$ represents the magnitude of received signal power. Hence,
we can define the energy radiation efficiency as $\frac{{{{\bf w}^H}{{\bf C}_0}{\bf w}}}{{{{\bf w}^H}{\bf w}}}$. The maximal energy radiation efficiency can be immediately obtained as $\lambda_{\max}({\bf C}_0)$, which is the maximal eigenvalue of ${\bf C}_0$. Then, we further define the normalized energy radiation efficiency as
\begin{align}
\eta = \frac{  \frac{ {\bf w}^H {\bf C}_0{\bf w} }{ {\bf w}^H {\bf w}  }  }{
\lambda_{\max}({\bf C}_0)
} ,
\end{align}
which is the ratio of the radiation efficiency to the maximum possible radiation efficiency and represents the level of maintained relative radiation efficiency.  The normalized radiation efficiency equals $2\times10^{-10}$ in AW-Mini-DS for the above example in Fig. 2, which means almost all of the radiation power has leaked out.

Based on the above observations, we propose a new antenna weighting scheme which targets at minimizing Doppler spread as well as maintaining the radiation efficiency.
Given a minimal required normalized radiation efficiency $\varepsilon$, the optimization problem can be formulated as:
\begin{align}\label{equ2}
{\hat{\bf w}}  = & \arg \mathop {\min }\limits_{\tilde{\bf w}}\frac{{{{\tilde{\bf w}}^H}{{\bf C}_2}{\tilde{\bf w}}}}{{{{\tilde{\bf w}}^H}{{\bf C}_0}{\tilde{\bf w}}}},  \\
 s.t. \ &\frac{{{{\tilde{\bf w}}^H}{{\bf C}_0}{\tilde{\bf w}}}}{{{{\tilde{\bf w}}^H}{\tilde{\bf w}}}} \ge \varepsilon  \cdot {\lambda _{\max }}({{\bf C}_0}). \nonumber
\end{align}

We can transfer the  optimization problem (\ref{equ2}) into
\begin{subequations}\label{equ7}
\begin{equation}
{\hat{\bf w}}  =  \arg \mathop {\min }\limits_{\tilde{\bf w}} {{\tilde{\bf w}}^H}{{\bf C}_2}{\tilde{\bf w}},
\end{equation}
\begin{equation}
s.t. \ {{\tilde{\bf w}}^H}{\tilde{\bf w}} \le \frac{1}{{\varepsilon  \cdot {\lambda _{\max }}({{\bf C}_0})}},
\hspace{-3em}
\end{equation}
\begin{equation}\label{equ1}
{{\tilde{\bf w}}^H}{{\bf C}_0}{\tilde{\bf w}} \ge 1.
\hspace{-1em}
\end{equation}
\end{subequations}

The detailed proof is omitted here due to the space limitation.
Note that (\ref{equ1}) is a non-convex constraint. We then resort to the
the SPCA method~\cite{Beck10} to solve the above non-convex problem iteratively. Specifically, given the optimal solution in the $l\rm{th}$ iteration $\hat{\bf w}_l$, we know ${{\tilde{\bf w}}^H}{{\bf C}_0}{\tilde{\bf w}} $ is lower bounded by its first Taylor expansion
 $f(\tilde{\bf w},{\hat{\bf w}}_l)$, where
\begin{align}
f(\tilde{\bf w},{\hat{\bf w}}_l) =
\hat{\bf w}_l^H {\bf C}_0 \hat{\bf w}_l +
2\Re\Big(
\textrm{Tr}\big( \hat{\bf w}_l^H {\bf C}_0 ( \tilde{\bf w} - \hat{\bf w}_l )  \big)
\Big).
\end{align}

Hence, the feasible set of $f(\tilde{\bf w},{\hat{\bf w}}_l) \ge 1$ serves as a subset of the original constraint ${{\tilde{\bf w}}^H}{{\bf C}_0}{\tilde{\bf w}} \ge 1$.
Then, the optimal solution in the $(l+1)\rm{th}$ iteration can be obtained by solving the following convex problem:
\begin{align}
{\hat{\bf w}}_{l+1} = &  \arg \mathop {\min }\limits_{\tilde{\bf w}} {{\tilde{\bf w}}^H}{{\bf C}_2}{\tilde{\bf w}}, \\
s.t. & \ {{\tilde{\bf w}}^H}{\tilde{\bf w}} \le \frac{1}{{\varepsilon  \cdot {\lambda _{\max }}({{\bf C}_0})}}, \nonumber \\
& 1 - f(\tilde{\bf w},{\hat{\bf w}}_l) \le 0.  \nonumber
\end{align}

Note that the initial feasible solution $\hat{\bf w}_0$ can be found by the Iterative Feasibility Search Algorith (IFSA) algorithm~\cite{Cheng10}. In Fig. 2, we further include the radiation pattern of the proposed scheme with $\varepsilon=0.5$ for comparison. As expected, it is observed that the sidelobe level of the proposed scheme is much lower than that of AW-Mini-DS and the mainlobe level of the proposed scheme is much higher than that of AW-Mini-DS.
This intuitively implies that the proposed method should provide much better radiation efficiency than AW-Mini-DS. More exactly,  the normalized radiation efficiencies of the AW-Mini-DS and the proposed scheme in this example are $2\times10^{-10}$ and $0.5$, respectively. In comparison, the corresponding normalized Doppler spreads are approximately given by $0.010$ and $0.012$, respectively. This verifies that the proposed scheme indeed could maintain radiation efficiency with slight sacrifice of Doppler spread suppression capability.

\section{Antenna Selection for Doppler Spread Suppression}

The massive MIMO system may suffer from high cost and complexity due to a large number of RF chains, making antenna selection urgently needed to reduce system complexity. In this section, we further exploit the idea of antenna selection for radiation efficiency aware Doppler spread suppression.

We consider that the $M$-element ULA of HST  has at most $N$ RF chains ($N\le M$). This is equivalent to impose an $l_0$-norm  constraint on the weighting vector, that is $\|{\bf w}\|_0 \le N$. Here, $l_0$-norm $|\cdot|_0$ returns the number of non-zero elements of given argument.
  Especially, when $N \ll M$, the constraint of $\|{\bf w}\|_0 \le N$ implies that ${\bf w}$ is a sparse vector. Then, the  antenna selection problem for radiation efficiency aware Doppler spread suppression can be formulated as:
\begin{subequations}\label{equ3}
\begin{equation}
{\hat{\bf w}}  =  \arg \mathop {\min }\limits_{\tilde{\bf w}}\frac{{{{\tilde{\bf w}}^H}{{\bf C}_2}{\tilde{\bf w}}}}{{{{\tilde{\bf w}}^H}{{\bf C}_0}{\tilde{\bf w}}}},
\hspace{1em}
\end{equation}
\begin{equation}
s.t. \ \frac{{{{\tilde{\bf w}}^H}{{\bf C}_0}{\tilde{\bf w}}}}{{{{\tilde{\bf w}}^H}{\tilde{\bf w}}}} \ge \varepsilon  \cdot {\lambda _{\max }}\big( {\bf C}_0(\tilde{\bf w}) \big),
\hspace{-4em}
\end{equation}
\begin{equation}
||{\tilde{\bf w}}|{|_0} \le  N.
\hspace{2em}
\end{equation}
\end{subequations}
Here, ${\bf C}_0(\tilde{\bf w})$ is a submatrix of ${\bf C}_0$
by deleting the column and row vectors corresponding to zero entries of trial weighting vector $\tilde{\bf w}$.

\begin{table}
\caption{\textbf{Algorithm:} Proposed antenna selection algorithm:}
\begin{tabular}{p{0.9\columnwidth}}
\hline
\textbf{Initialization:}  set ${\sigma _l} = {\sigma_{\min }}$ and ${\sigma_h} = {\sigma_{\max }}$\\
\textbf{Repeat:}\\
$1:$ Set $\sigma = \frac{\sigma_l + \sigma_h}{2}$; \\
$2:$ Solve the problem (\ref{equ8}) by exploiting SPCA method;  \\
\quad \quad if $\| \hat{\bf  w } \|_0 > N$, set ${\sigma _l} = \sigma$. \\
\quad \quad  if $ \| \hat{\bf w } \|_0 < N$, set ${\sigma _h} = \sigma$. \\
\quad \quad  if $\|\hat{\bf w } \|_0 = N $, break. \\
\textbf{Return:} The non-zero elements of $\hat{\bf w } $ correspond to the selected antennas. \\
\hline
\end{tabular}
\end{table}

Bearing in mind that ${\bf C}_0(\tilde{\bf w})$ is parameterized by $\tilde{\bf w}$, directly solving the above problem (\ref{equ3}) should be a nontrivial task. Next, we propose a two-step sub-optimal solution to the problem. The first step is to optimize the antenna selection for Doppler spread minimization without consideration of radiation efficiency. Once the antenna selection is determined, the radiation efficiency aware Doppler spread suppression scheme proposed in Section III can be immediately employed in the second step. Thus, in the following we will focus in the first step, which can be formulated as:
\begin{align} \label{equ4}
{\hat{\bf w}}  =  & \arg \mathop {\min }\limits_{\tilde{\bf w}}\frac{{{{\tilde{\bf w}}^H}{{\bf C}_2}{\tilde{\bf w}}}}{{{{\tilde{\bf w}}^H}{{\bf C}_0}{\tilde{\bf w}}}}, \nonumber\\
s.t. \ & ||{\tilde{\bf w}}|{|_0} \le  N.
\end{align}

To circumvent the intractable non-convex problem (\ref{equ4}),
we formulate the following function $S(\sigma)$:
\begin{align}\label{equ5}
S(\sigma) = & \min_{\tilde{\bf w}} \|\tilde{\bf w}\|_0      \nonumber \\
s.t. \ & \frac{{{{\tilde{\bf w}}^H}{{\bf C}_2}{\tilde{\bf w}}}}{{{{\tilde{\bf w}}^H}{{\bf C}_0}{\tilde{\bf w}}}} \le \sigma.
\end{align}

The above problem (\ref{equ5}) tries to find the sparsest $\tilde{\bf w}$, given a maximal allowable Doppler spread $\sigma$.
Moreover, the function $S(\sigma)$ is a decreasing function of the allowable Doppler spread $\sigma$. It can be observed that, the problem of (\ref{equ4}) is equivalent to solve the equation $S(\sigma) = N$. This can be solved efficiently via bisection method from an interval $[\sigma_{\min}, \sigma_{\max}]$ with $S(\sigma_{\min}) \ge N $ and  $S(\sigma_{\max}) \le N$. Here, $\sigma_{\min}$ and $\sigma_{\max}$ can be set as the minimal Doppler spreads via AW-Mini-DS scheme with a fully connected $M$-element and $N$-element ULA, respectively.

Next, given a fixed $\sigma$,  in order to solve the problem (\ref{equ5}), a good approach is to approximate the objective function by the well-known $l_1$-norm~\cite{Nai10}, that is
\begin{align}\label{equ6}
 \hat{\bf w} = &  \arg \mathop {\min }\limits_{\tilde{\bf w}}||{\tilde{\bf w}}|{|_1},\nonumber \\
s.t. \ & \frac{{{{\tilde{\bf w}}^H}{{\bf C}_2}{\tilde{\bf w}}}}{{{{\tilde{\bf w}}^H}{{\bf C}_0}{\tilde{\bf w}}}} \le \sigma.
\end{align}

The above problem (\ref{equ6}) can be further transformed into
\begin{align}\label{equ8}
  \hat{\bf w} =  & \min_{\tilde{\bf w}} \| {\bf w} \|_1 \\
 s.t. \quad  & \tilde{\bf w}^H {\bf C}_2 \tilde{\bf w} \le \sigma,  \nonumber \\
 & \tilde{\bf w}^H {\bf C}_0 \tilde{\bf w} \ge 1. \nonumber
\end{align}

Similar to (\ref{equ7}), the non-convex problem (\ref{equ8}) can be iteratively solved by with the aid of SPCA method.
In  summary, the proposed antenna selection scheme in the first step can be described in table I.

\section{Simulation Results}
\label{sec:Simulation Results}

In this section, we present simulation results to verify the proposed studies. We consider a ULA with a total of $M=64$ antennas and normalized antenna spacing of 0.45.

\begin{figure}[t]
\begin{center}
\includegraphics[width=85mm]{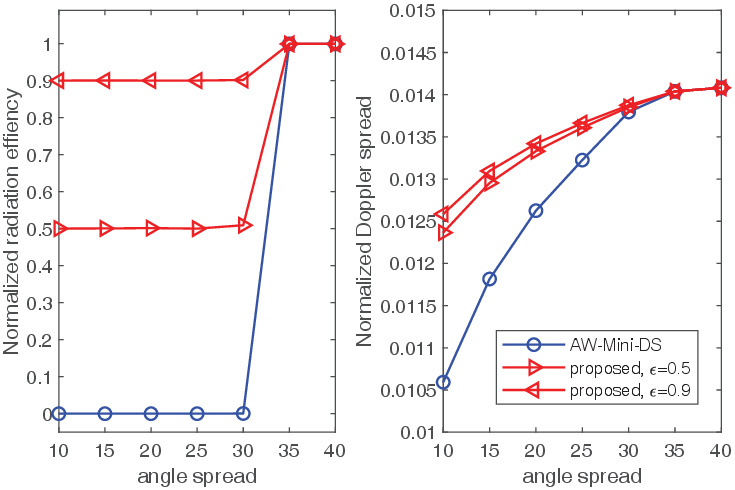}
\vspace{-5mm}
\end{center}
\caption{  Performance comparison between AW-Mini-DS and the proposed method in terms of both normalized radiation efficiency and Doppler spread. }
\end{figure}

In the first example, we consider the case of fully connected ULA. We compare the performance between AW-Mini-DS and the proposed scheme in terms of both normalized radiation efficiency and Doppler spread as the angle spread $\theta_R - \theta_L$ increases in Fig. 3.  The center AOD is fixed as $(\theta_L + \theta_R)/2 = 90^\circ$.
We can make the following observations: As the angle spread decreases, AW-Mini-DS can always provide the lowest Doppler spread (the best Doppler spread suppression). Unfortunately,  the radiation efficiency of AW-Mini-DS quickly drops almost to  zero once the angle spread becomes below $30^\circ$. In comparison,  the results verify the effectiveness of the proposed radiation efficiency aware scheme. It is seen that, the proposed scheme could maintain the radiation efficiency according to the constraint $\varepsilon$ at the expense of slight Doppler spread suppression capability.  Take the angle spread of $10^\circ$ as an example.
As compared to AW-Mini-DS, the proposed scheme  raises  the Doppler spread by a factor of round $20\%$, but in the meanwhile improves the radiation efficiency from almost zero to 0.9. Hence, the proposed scheme has provided more balanced performance as compared to AW-Mini-DS.

Next, we consider the case of partly connected ULA, where $M$-element ULA has at most $N$ RF chains. We display the evolution of normalized Doppler spread of the proposed antenna selection scheme as the number of RF chains increases in Fig. 4. We assume $\theta_L = 30^{\circ}$ and $\theta_R = 60^{\circ}$. For comparison, we also include the results of AW-Mini-DS in this figure.
We consider that AW-Mini-DS employs the fully connected ULA with $N$ and $M$ RF chains, respectively. As expected, the AW-Mini-DS scheme with $N$ and $M$ RF chains serve as the upper and lower bound of the proposed scheme, respectively. With the same number of RF chains, the proposed scheme can greatly reduce the Doppler spread as compared to AW-Mini-DS.
Moreover, as the number of RF chains increases, the proposed scheme could closely approach the lower bound, which demonstrates the effectiveness of the proposed antenna selection scheme.

\begin{figure}[t]
\begin{center}
\includegraphics[width=70mm]{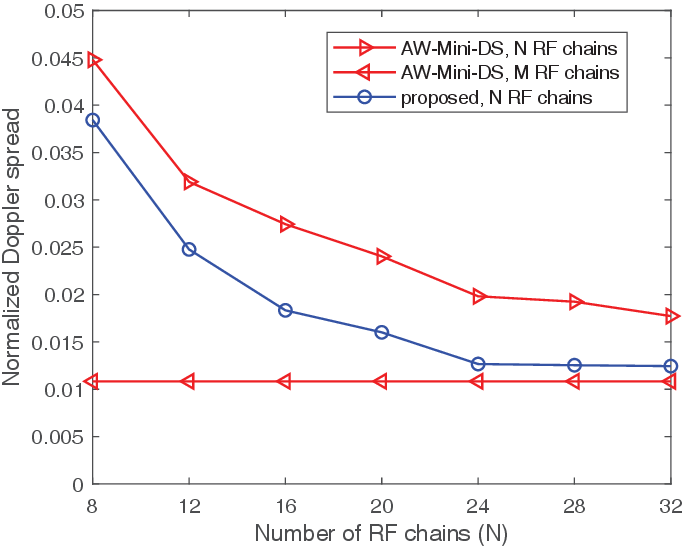}
\vspace{-5mm}
\end{center}
\caption{  Normalized Doppler spread of the proposed scheme as the number of RF chains increases. }
\vspace{-5mm}
\end{figure}

\section{Conclusions}

In this paper, we considered the angle-domain Doppler shifts compensation scheme with maintaining radiation efficiency for high-mobility uplink communication. The idea of antenna selection is further exploited to reduce the number of required RF chains. Numerical results were provided to corroborate the proposed studies.

\end{document}